\documentclass[a4paper,12pt,reqno]{amsart}

\usepackage{pgfplots}
\usepackage{graphics,amssymb,tikz}
\usetikzlibrary{arrows}
\usepackage{amsfonts,amsmath,latexsym,hyperref}
\usepackage{amsthm}
\usepackage{color}
\usepackage{fullpage}
\usepackage{booktabs}

\newcommand {\Z}{\mathbb{Z}}

\newcommand {\C}{\mathbb{C}}
\newcommand{\ep}{\varepsilon}
\newcommand{\VM}{V^{\rm S}_{\rm max}}
\newcommand{\VY}{V^{\rm T}_{\rm max}}
\newcommand{\VH}{V^{\rm H}}
\newcommand{\prodprime}{\sideset{}{'}\prod_{\lambda \in \Lambda}}
\newcommand{\sumprime}{\sideset{}{'}\sum_{\lambda \in \Lambda}}
\newtheorem {thm} {Theorem}

\newtheorem {lemma}{Lemma}

\newtheorem {rmk} {Remark}
\newtheorem* {claim} {Claim}
\newcommand{\beq}{\begin{equation}}
\newcommand{\eeq}{\end{equation}}
\numberwithin{equation}{section}

\begin {document}

\title[The maximum voltage drop]{The maximum voltage drop in an on-chip power distribution network: 
analysis of square, triangular and hexagonal power pad arrangements}
\author{Tom Carroll}
\address{Department of Mathematics\\
University College Cork\\
Cork, Ireland}
\email{t.carroll@ucc.ie}
\author{Joaquim Ortega-Cerd\`a }
\address{Departament de Matem\`atica Aplicada i An\`alisi\\
Universitat de Barcelona\\
Gran Via 585\\
08007 Barcelona, Spain.}
\email{jortega@ub.edu}

\date{\today}

\begin{abstract}
A mathematical model of the voltage drop which arises in on-chip 
power distribution networks is used to compare 
the maximum voltage drop in the case of different geometric arrangements of the 
pads supplying power to the chip. 
These include the square or Manhattan power pad arrangement 
which currently predominates,
as well as equilateral triangular and hexagonal arrangements.
In agreement with findings in the literature and with physical and SPICE models, 
the equilateral power pad arrangement, 
independent of the underlying power mesh configuration, is found to
minimize the maximum voltage drop. 
This headline finding is a consequence of relatively simple formulas for the 
voltage drop, with explicit error bounds, 
which are established using complex analysis techniques, 
and elliptic functions in particular. 
\end{abstract}

\maketitle


\section{Introduction}\label{sec1}
Control of the maximum voltage drop between power distribution pads is a 
factor of increasing importance in the design of the power distribution network
(PDN) of modern IC computer chips. 
The voltage drop between power pads
depends both on the current flowing in the power mesh between the pads 
and on the electrical resistance in the power mesh.  
The physical layout of a computer chip and the
interaction between the chip and its power distribution network are described in
detail by Shakeri and Meindl \cite{SM} in the context of both wire-bond and
flip-chip PDN design. They focus on the dominant paradigm in which
the power pads and the power mesh are arranged in a square grid, 
which is known as the Manhattan architecture, they 
derive the equations governing the voltage drop and provide the
leading terms of the solution. The Y-architecture, in which pads are arranged in
an equilateral lattice and the power mesh is also arranged in an equilateral
grid, is considered by Chen \emph{et al.}\ \cite{Y-arch}. Analytical and
simulation results are obtained which indicate a 5\% reduction in the 
maximum voltage drop in the case of a single layer Y-architecture compared to the
single layer Manhattan architecture. 

Aquareles \emph{et al.}\ \cite{Agua} put the mathematical aspects of the work
of Shakeri and Meindl on a firm footing. They obtain an asymptotic formula for
the maximum voltage drop in terms of the size of the pads, including higher
order terms that would seem to be beyond the techniques in \cite{SM}. The main
mathematical tool they use is that of matched asymptotic expansions. In the
present work, we use a complex analysis method to derive an expression for the
maximum voltage drop in the case of the square pad arrangement. This method
is simpler and more direct than the approach in \cite{Agua} and covers, without 
additional effort, the case of pads arranged in an equilateral triangular array.
With a little extra work, the method extends to treat the case of pads arranged in a
hexagonal pattern. 

The results we obtain suggest that the smaller maximum voltage drops 
observed by Chen \emph{et al.} in \cite{Y-arch} are due to the 
arrangement of the pads in an equilateral array and are independent 
of the configuration of the underlying power mesh. 
That is, for an equilateral disposition of the power pads superposed over a fine 
Manhattan power mesh there will be a similar voltage drop to that observed in the
Y-architecture.

We also obtain formulas for
the maximum voltage drop in each of these configurations (square, triangular
and hexagonal). It is
found than the hexagonal pad arrangement has the largest voltage drop of the
three configurations considered. Even so, it may also be useful to have an
explicit formula for the voltage drop in this case since, however important,
control of the maximum voltage drop is but one of several constraints in the
design of an on-chip PDN. Finally, the availability of explicit formulas makes
it possible to accurately predict the maximum voltage drop at an early point in
the circuit design stage, thereby obviating the need for costly re-design.
%
%
\section{Mathematical model of the voltage drop}\label{sec2}
In this section we describe the mathematical model 
of the power distribution network and  the associated voltage drop as derived by Shakeri and Meindl \cite{SM}.

The surface of the integrated circuit is modeled as an infinite complex 
plane in which the power pads 
of the power distribution network are modeled as circular disks. 
Power to the chip is supplied through these power pads and 
distributed through a fine grid of wires called the power mesh. 
The square and triangular arrangement of the pads are displayed below.
The planar region consisting of the complex plane with
these circular disks removed is denoted by $\Omega$.
Under the assumption of uniform current flow between pads, the voltage
drop satisfies the equation $\Delta u=c$ as the power mesh (triangular
or square) gets finer.

The constant $c$ on the right hand side of this partial differential equation
codes for the resistance properties of the wires of the mesh and the current
drawn from the power network. In order to make a fair comparison between the
voltage drop across different power distribution network configurations, the
resistance properties of the underlying integrated circuits (IC) and the current
drawn need to be the same, that is we need to use the same constant $c$ in all
cases. Moreover, since we measure the relative change in the maximum voltage
drop across different arrangements of the power pads, and since the solution to
$\Delta u = c$ is proportional to $c$, it suffices to take the common value
$c=1$ in the modeling equation. Next, the power distribution pads are held at
constant voltage, which we may take to equal zero. Thus the governing partial
differential equation for the voltage in the region $\Omega$ between the power
pads is  
\beq
\label{upde}
  \left\{
  \begin{aligned}
  \Delta u  = 1 & \mbox{ in } \Omega,  \\
  u = 0  & \mbox{ on } \partial \Omega. 
  \end{aligned}
  \right.
\eeq
The voltage between the pads will then be negative since $u$ is subharmonic and
the pads themselves are held at voltage 0, 
while the voltage drop relative to the pads will simply be $-u$.
It is interesting to note that the solution of the partial differential equation
$\Delta u =-2$ in a domain $D$, also with zero Dirichlet boundary conditions,
describes the expected exit time of Brownian motion from the domain. Thus, the
problem of determining the maximum voltage drop is mathematically equivalent to
determining the maximum expected lifetime of Brownian motion in the domain
complementary to the power pads. 
\definecolor{dcrutc}{rgb}{0.86,0.08,0.24}
\definecolor{ffqqqq}{rgb}{1,0,0}
\definecolor{uququq}{rgb}{0.25,0.25,0.25}
\definecolor{qqqqff}{rgb}{0,0,1}
%
%
\newsavebox{\mysquare}
\savebox{\mysquare}{
\tikz
{\draw[color=qqqqff]  (-0.7,-0.7) rectangle (0.7,0.7);
\fill [color=uququq]  (-0.7,-0.7) circle (3.0pt);
\fill [color=uququq]  (0.7,0.7) circle (3.0pt);
\fill [color=uququq]  (-0.7,0.7) circle (3.0pt);
\fill [color=uququq]  (0.7,-0.7) circle (3.0pt);}
}
\newsavebox{\myredsquare}
\savebox{\myredsquare}{
\tikz
{\draw[color=red,dashed]  (-0.7,-0.7) rectangle (0.7,0.7);}
}
\newsavebox{\mytriangle}
\savebox{\mytriangle}{
\tikz
{\draw[color=qqqqff]  (-0.7522,-0.43428) --  (0.7522,-0.43428) -- (0,0.86857) -- (-0.7522,-0.43428);
\fill [color=uququq]  (-0.7522,-0.43428) circle (3.0pt);
\fill [color=uququq]  (0.7522,-0.43428) circle (3.0pt);
\fill [color=uququq]  (0,0.86857) circle (3.0pt);}
}
\newsavebox{\myrotatedtriangle}
\savebox{\myrotatedtriangle}{
\tikz
{\draw[color=qqqqff]  (-0.7522,0.86857) --  (0.7522,0.86857) -- (0,-0.43428) -- (-0.7522,0.86857);
\fill [color=uququq]  (-0.7522,0.86857) circle (3.0pt);
\fill [color=uququq]  (0.7522,0.86857) circle (3.0pt);
\fill [color=uququq]  (0,-0.43428) circle (3.0pt);}
}
\newsavebox{\myreddiamond}
\savebox{\myreddiamond}{
\tikz
{\draw[color=red, dashed]  (-0.7522,0) -- (0,1.30285) -- (0.7522,0) -- (0,-1.30285) -- (-0.7522,0);}
}
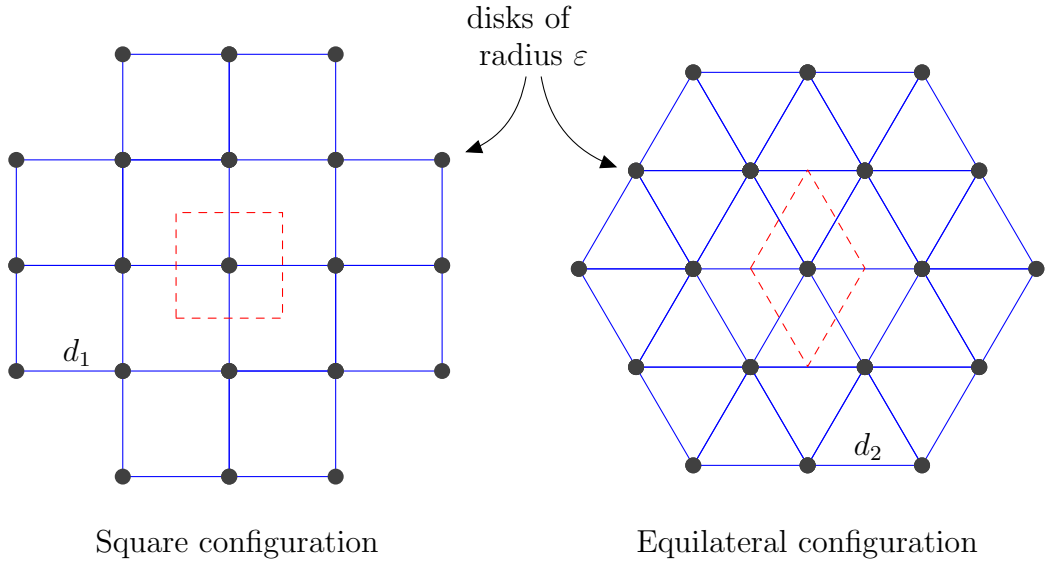
\begin{figure}
\caption{Square and equilateral arrangement of pads}\label{fig1}
\begin{center}
\begin{tikzpicture}[line cap=round,line join=round,>=triangle 45,x=1.0cm,y=1.0cm]
\draw (-2.8,1.4) node {\usebox{\mysquare}};
\draw (-1.4,0) node {\usebox{\mysquare}};
\draw (0,0) node {\usebox{\mysquare}};
\draw (0,1.4) node {\usebox{\mysquare}};
\draw (-4.2,0) node {\usebox{\mysquare}};
\draw (-4.2,1.4) node {\usebox{\mysquare}};
\draw (-2.8,2.8) node {\usebox{\mysquare}};
\draw (-1.4,2.8) node {\usebox{\mysquare}};
\draw (-2.8,-1.4) node {\usebox{\mysquare}};
\draw (-1.4,-1.4) node {\usebox{\mysquare}};
\node at (-2,-3) {Square configuration};
\draw (-2.1,0.7) node {\usebox{\myredsquare}};
\draw(4,0) node {\usebox{\mytriangle}};
\draw(5.5044,0) node {\usebox{\mytriangle}};
\draw(7.0088,0) node {\usebox{\mytriangle}};
\draw(4.7522,1.30285) node {\usebox{\mytriangle}};
\draw(6.2566,1.30285) node {\usebox{\mytriangle}};
\draw(4,1.30285) node {\usebox{\myrotatedtriangle}};
\draw(5.5044,1.30285) node {\usebox{\myrotatedtriangle}};
\draw(7.0088,1.30285) node {\usebox{\myrotatedtriangle}};
\draw(3.2478,1.30285) node {\usebox{\mytriangle}};
\draw(3.2478,0) node {\usebox{\myrotatedtriangle}};
\draw(7.76,1.30285) node {\usebox{\mytriangle}};
\draw(7.76,0) node {\usebox{\myrotatedtriangle}};
\draw(4,2.6059) node {\usebox{\mytriangle}};
\draw(5.5044,2.6059) node {\usebox{\mytriangle}};
\draw(7.0088,2.6059) node {\usebox{\mytriangle}};
\draw(4.7522,2.6059) node {\usebox{\myrotatedtriangle}};
\draw(6.2566,2.6059) node {\usebox{\myrotatedtriangle}};
\draw(4,-1.30285) node {\usebox{\myrotatedtriangle}};
\draw(5.5044,-1.30285) node {\usebox{\myrotatedtriangle}};
\draw(7.0088,-1.30285) node {\usebox{\myrotatedtriangle}};
\draw(4.7522,-1.30285) node {\usebox{\mytriangle}};
\draw(6.2566,-1.30285) node {\usebox{\mytriangle}};
\node at (5.5,-3) {Equilateral configuration};
\draw(5.5044,0.66) node {\usebox{\myreddiamond}};
\node at (1.7,4) {disks of};
\node at (1.9, 3.5) {radius $\varepsilon$};
\draw[->](1.8,3.2) to [out = -100, in = 20] (1,2.2);
\draw[->](2,3.2) to [out = -80, in = 160] (3,2);
\node at (-4.1,-0.45) {$d_1$};
\node at (6.3,-1.7) {$d_2$};
\end{tikzpicture}
\end{center}
\end{figure}

The partial differential equation \eqref{upde} obeys a scaling law: If $u(z)$ is
the solution of $\Delta u=1$ in a domain $D$ then $v(w) = r^2 u(w/r)$ is the
solution of $\Delta v = 1$ in the domain $rD$. Thus, if the radius of the power
pads and the spacing between their centres both change by a factor of $r$ then
the maximum voltage drop changes by a factor $r^2$. If we know the voltage drop
for all values of the radius of the power pads for some \emph{fixed} spacing
between their centres then we can scale this result to determine the voltage
drop in the case of any power pad radius and any spacing between their centres. 

Next, in order to make a fair comparison between different geometric power pad
configurations, the proportion of the area on the chip occupied by
the power pads (let's call it $p$) should be the same in each case. Notice that
$p$, the area of the power pads per unit area on the chip, does not change under
the scaling $z\to rz$ discussed above, whereas the voltage drop changes by a
factor $r^2$. Thus, even for prescribed areal density $p$ of the power pads, the
voltage drop can be made as small as one wishes by taking smaller pads closer
together. Thus, in order to make a fair comparison between different
configurations, it is necessary not only to ensure that the aereal density of
the power pads are the same in each configuration but also to specify the radius
$\varepsilon$ of each pad. The values of $p$ and $\epsilon$ then determine the
spacing between the pads. (Alternatively, one could instead specify the spacing
between the pads rather than their radius, but this seems less natural.) 

Referring to Figure~\ref{fig1}, each pad in the square arrangement configuration
lies at the centre of a square of side $d_1$ which does not overlap with the
corresponding square for any other pad. Thus the aereal density $p =
\pi\varepsilon^2/d_1^2$ in this case. 
For an equilateral triangular arrangement, 
each pad lies at the centre of a diamond of area
$\sqrt{3}\,d_2^2/2$ which does not overlap with the corresponding diamond for
any other pad. Thus the aereal density of the pads in the
equilateral configuration is $p=2\pi\varepsilon^2/(\sqrt{3}\,d_2^2)$. For
prescribed common radius $\varepsilon$ of the pads, the aereal density of the
pads will be the same in both configurations once 
\beq\label{2.2}
d_2^2 = \frac{2}{\sqrt{3}}\,d_1^2.
\eeq
Assuming, therefore, that in the square arrangement we have power pads of radius
$\varepsilon$ whose centres are unit distance apart, in the equilateral
arrangement we should have power pads of radius $\varepsilon$ whose centres are
$d_2 = \sqrt{2}/\root 4 \of 3 \simeq 1.0745$ apart. 

\begin{figure}
\caption{Hexagonal configuration}\label{fig2}
\begin{center}
\begin{tikzpicture}[line cap=round,line join=round,>=triangle 45,x=1.0cm,y=1.0cm]
\draw [color=qqqqff] (2,0)-- (3,0);
\draw [color=qqqqff] (3,0)-- (3.5,0.87);
\draw [color=qqqqff] (3.5,0.87)-- (3,1.73);
\draw [color=qqqqff] (3,1.73)-- (2,1.73);
\draw [color=qqqqff] (2,1.73)-- (1.5,0.87);
\draw [color=qqqqff] (1.5,0.87)-- (2,0);
\draw [color=qqqqff] (2,1.73)-- (3,1.73);
\draw [color=qqqqff] (3,1.73)-- (3.5,2.6);
\draw [color=qqqqff] (3.5,2.6)-- (3,3.46);
\draw [color=qqqqff] (3,3.46)-- (2,3.46);
\draw [color=qqqqff] (2,3.46)-- (1.5,2.6);
\draw [color=qqqqff] (1.5,2.6)-- (2,1.73);
\draw [color=qqqqff] (3,1.73)-- (3.5,0.87);
\draw [color=qqqqff] (3.5,0.87)-- (4.5,0.87);
\draw [color=qqqqff] (4.5,0.87)-- (5,1.73);
\draw [color=qqqqff] (5,1.73)-- (4.5,2.6);
\draw [color=qqqqff] (4.5,2.6)-- (3.5,2.6);
\draw [color=qqqqff] (3.5,2.6)-- (3,1.73);
\draw [color=qqqqff] (3.5,0.87)-- (3,0);
\draw [color=qqqqff] (3,0)-- (3.5,-0.87);
\draw [color=qqqqff] (3.5,-0.87)-- (4.5,-0.87);
\draw [color=qqqqff] (4.5,-0.87)-- (5,0);
\draw [color=qqqqff] (5,0)-- (4.5,0.87);
\draw [color=qqqqff] (4.5,0.87)-- (3.5,0.87);
\draw [color=qqqqff] (3,0)-- (2,0);
\draw [color=qqqqff] (2,0)-- (1.5,-0.87);
\draw [color=qqqqff] (1.5,-0.87)-- (2,-1.73);
\draw [color=qqqqff] (2,-1.73)-- (3,-1.73);
\draw [color=qqqqff] (3,-1.73)-- (3.5,-0.87);
\draw [color=qqqqff] (3.5,-0.87)-- (3,0);
\draw [color=qqqqff] (1.5,-0.87)-- (2,0);
\draw [color=qqqqff] (2,0)-- (1.5,0.87);
\draw [color=qqqqff] (1.5,0.87)-- (0.5,0.87);
\draw [color=qqqqff] (0.5,0.87)-- (0,0);
\draw [color=qqqqff] (0,0)-- (0.5,-0.87);
\draw [color=qqqqff] (0.5,-0.87)-- (1.5,-0.87);
\draw [color=qqqqff] (1.5,0.87)-- (2,1.73);
\draw [color=qqqqff] (2,1.73)-- (1.5,2.6);
\draw [color=qqqqff] (1.5,2.6)-- (0.5,2.6);
\draw [color=qqqqff] (0.5,2.6)-- (0,1.73);
\draw [color=qqqqff] (0,1.73)-- (0.5,0.87);
\draw [color=qqqqff] (0.5,0.87)-- (1.5,0.87);
\draw [dashed,color=red] (1,1.73)-- (2.5,0.87);
\draw [dashed,color=red] (2.5,2.6)-- (2.5,0.87);
\draw [dashed,color=red] (4,1.73)-- (2.5,0.87);
\draw [dashed,color=red]  (4,0)-- (2.5,0.87);
\draw [dashed,color=red]  (2.5,-0.87)-- (2.5,0.87);
\draw [dashed,color=red]  (1,0)-- (2.5,0.87);
\draw [dashed,color=red]  (1,0)-- (2.5,-0.87);
\draw [dashed,color=red]  (2.5,-0.87)-- (4,0);
\draw [dashed,color=red]  (4,0)-- (4,1.73);
\draw [dashed,color=red]  (4,1.73)-- (2.5,2.6);
\draw [dashed,color=red]  (2.5,2.6)-- (1,1.73);
\draw [dashed,color=red]  (1,1.73)-- (1,0);
\fill [color=uququq] (2,0) circle (3.0pt);
\fill [color=uququq] (3,0) circle (3.0pt);
\fill [color=uququq] (3.5,0.87) circle (3.0pt);
\fill [color=uququq] (3,1.73) circle (3.0pt);
\fill [color=uququq] (2,1.73) circle (3.0pt);
\fill [color=uququq] (1.5,0.87) circle (3pt);
\fill [color=uququq] (3.5,2.6) circle (3.0pt);
\fill [color=uququq] (3,3.46) circle (3.0pt);
\fill [color=uququq] (2,3.46) circle (3.0pt);
\fill [color=uququq] (1.5,2.6) circle (3.0pt);
\fill [color=uququq] (4.5,0.87) circle (3.0pt);
\fill [color=uququq] (5,1.73) circle (3.0pt);
\fill [color=uququq] (4.5,2.6) circle (3.0pt);
\fill [color=uququq] (3.5,2.6) circle (3.0pt);
\fill [color=uququq] (3.5,-0.87) circle (3.0pt);
\fill [color=uququq] (4.5,-0.87) circle (3.0pt);
\fill [color=uququq] (5,0) circle (3.0pt);
\fill [color=uququq] (4.5,0.87) circle (3.0pt);
\fill [color=uququq] (1.5,-0.87) circle (3.0pt);
\fill [color=uququq] (2,-1.73) circle (3.0pt);
\fill [color=uququq] (3,-1.73) circle (3.0pt);
\fill [color=uququq] (3.5,-0.87) circle (3.0pt);
\fill [color=uququq] (1.5,0.87) circle (3.0pt);
\fill [color=uququq] (0.5,0.87) circle (3.0pt);
\fill [color=uququq] (0,0) circle (3.0pt);
\fill [color=uququq] (0.5,-0.87) circle (3.0pt);
\fill [color=uququq] (1.5,2.6) circle (3.0pt);
\fill [color=uququq] (0.5,2.6) circle (3.0pt);
\fill [color=uququq] (0,1.73) circle (3.0pt);
\fill [color=uququq] (0.5,0.87) circle (3.0pt);
\fill [color=red!50!white] (1,1.73) circle (3pt);
\fill [color=red!50!white] (2.5,2.6) circle (3pt);
\fill [color=red!50!white] (4,1.73) circle (3pt);
\fill [color=red!50!white] (4,0) circle (3pt);
\fill [color=red!50!white] (2.5,0.87) circle (3pt);
\draw[color=uququq, anchor =  west] (2.61,0.99) node {0};
\fill [color=red!50!white] (1,0) circle (3pt);
\fill [color=red!50!white] (2.5,-0.87) circle (3pt);
\fill [color=red!50!white] (4,0) circle (3pt);
\fill [color=red!50!white] (4,1.73) circle (3pt);
\fill [color=red!50!white] (2.5,2.6) circle (3pt);
\fill [color=red!50!white] (1,1.73) circle (3pt);
\node at (4.98,0.57) {$d_3$};
\draw[->](3,2.3) to [out = 75, in = 150] (4.9,3);
\node at (5.4,3) {$\sqrt{3}\,d_3$};
\end{tikzpicture}
\end{center}
\end{figure}
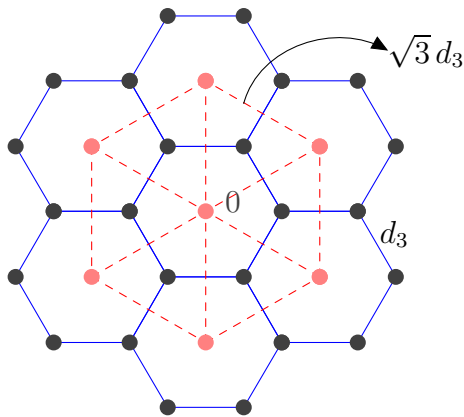

In the case of the hexagonal configuration, as shown in Figure~\ref{fig2}, each
pad lies at the centre of an equilateral triangle of sidelength $\sqrt{3}\,d_3$
which does not overlap with the corresponding triangle for any other pad. The
area of this triangle is $\sqrt{3}(\sqrt{3}\,d_3)^2/4 = 3\sqrt{3}\,d_3^2/4$, so
that the aereal density for the hexagonal configuration is $p =
4\pi\varepsilon^2 / ( 3\sqrt{3}\, d_3^2)$. In order that this agrees with the
aereal density $p=\pi \varepsilon^2$ for the previous configurations, 
we need
\[
d_3 = \frac{2}{\root 4 \of {27}}.
\]
In this case, each hexagon has area 2. 

%
%
\section{Main numerical results}
Analytic formulas for the voltage drop in the case of each of the arrangements
of the pads considered above are established in Sections~\ref{MY} and \ref{H}.
These yield the following bounds for the maximum voltage drop. In terms of the
radius $\varepsilon$ of the pads, the maximum voltage drop $\VM$ in the case of
the square arrangement is 
\beq\label{VM}
\VM(\varepsilon) = \frac 1{2\pi} \log \frac1\varepsilon 
 	- 0.153418893205  + \frac 14 \varepsilon^2 + O(\varepsilon^3)
\eeq
The maximum voltage drop $\VY$ in the case of the triangular configuration is 
\beq\label{VY}
\VY(\varepsilon)=\frac 1{2\pi} \log \frac
1\varepsilon - 0.166549975068  + \frac 14 \varepsilon^2 + O(\varepsilon^6). 
\eeq
In the case of the hexagonal configuration, the voltage drop at the centre of a hexagon
is 
\beq\label{VH}
\VH(\varepsilon)=\frac 1{2\pi} \log \frac 1\varepsilon -  0.111391075030 
+ \frac 14 \varepsilon^2 + O(\varepsilon^3)
\eeq
It is notable that, apart from the error term, the maximum voltage drop has the
same dependence on the pad size in all three cases, the only difference being in
the constant term. The conclusion is that the hexagonal pad arrangement has the
worst voltage drop among the configurations which we consider, the best being
the triangular lattice with the standard square lattice being in an intermediate
position.

One intuitive explanation of this situation is that though in the hexagonal arrangement
there are six disks around the origin they are, crucially, further separated
from the origin than in the other configurations considered. 
It is possible to fit a bigger disk around the origin which does not meet 
the boundary of $\Omega$ and this allows the Brownian motion to 
increase its expected lifespan.

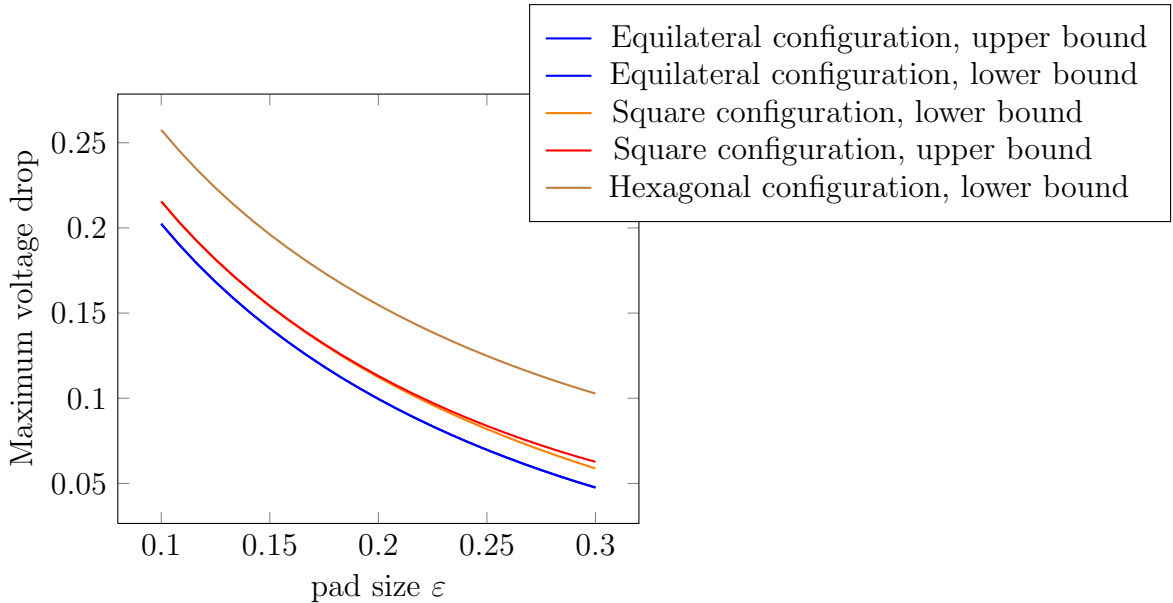
\begin{figure}\label{fig3}
\caption{Graphs of maximum voltage drop with varying pad size}
\begin{center}
\begin{tikzpicture}
    \pgfkeys{/pgf/number format/.cd,fixed,precision=2,use period}
    \tikzstyle{every axis legend}+=[
            anchor=north west, outer sep=-1.3cm, inner sep=0.2cm]
    \begin{axis}[
        xlabel= {pad size $\varepsilon$},
        ylabel={Maximum voltage drop},
        every axis plot post/.style=  thick,
    ]
    \addplot [blue][domain = 0.1:0.3] {-0.166549975068 - ln(x)/(2*pi) +0.25*x^2
    +0.69020942/(2*pi)*(x^6+x^12)};
    \addlegendentry{\ Equilateral configuration, upper bound\hfill}
    \addplot [blue][domain = 0.1:0.3] {-0.166549975068 - ln(x)/(2*pi) +0.25*x^2
    -0.69020942/(2*pi)*(x^6+x^12)};
    \addlegendentry{\ Equilateral configuration, lower bound\ \  }
    \addplot [orange] [domain = 0.1:0.3] {-0.153418893205 - ln(x)/(2*pi) 
+0.25*x^2-1.50670300/(2*pi)*(x^4+x^8)};
    \addlegendentry{\ Square configuration, lower bound \ \ \ \ \ \ }
    \addplot [red] [domain = 0.1:0.3] {-0.153418893205 - ln(x)/(2*pi) 
+0.25*x^2+1.50670300/(2*pi)*(x^4+x^8)};
    \addlegendentry{\ Square configuration, upper bound \ \ \ \ \ }
    \addplot [brown] [domain = 0.1:0.3] {-0.11139107503025511303 - ln(x)/(2*pi) 
   +0.25*x^2};
    \addlegendentry{\ Hexagonal configuration, lower bound\ \ \ }
    \end{axis}
\end{tikzpicture}
\end{center}
\end{figure}

The analytical results to come in Sections~\ref{MY} and \ref{H} yield explicit
error bounds in \eqref{VM}, \eqref{VY} and \eqref{VH}, which are displayed
graphically in Figure~\ref{fig3}. The curves represent upper and lower bounds for 
the maximum voltage drops $\VM(\ep)$ and $\VY(\ep)$ 
which take account of the error terms. 
The graph in the hexagonal case shows the voltage drop $\VH(\ep)$ 
at the centre of a hexagon. Presumably, this \emph{is} the maximum 
voltage drop, that is the maximum voltage drop presumably occurs at the centre 
of a hexagon,  but in any case the maximum voltage drop is at least this large. 
Thus, even at the limits of the error bounds, the
equilateral arrangement outperforms the square and the hexagonal arrangements
for all pad sizes. Note also that the error bounds are seen to be quite tight 
in both the square and the equilateral configurations, so that the formulas 
\eqref{VM} and \eqref{VY} are accurate. 
Note that the range of pad size $\ep$ (from $0.1$ to $0.3$)
relative to the distance between the distance between the centres of the pads
($d_1$, $d_2$, $d_3$, each of which is about unit size) is informed by industry
norms (see \cite[Table~III]{SM}).

In order to test the robustness of these analytical results we assembled two 
boards, each with a rectangular mesh of resistances. A constant current sink
was connected at each node. In one of the boards the voltage distribution
was through a collection of pads in a square configuration and in the
other the pads were in a triangular configuration. 
All pads were held at 5V. 
The maximum voltage drop was measured for each board. 
It was 1.91V in the triangular pad setting versus 2.03V in the square setting. 
SPICE simulations with the same configuration gave voltage drops of 1.94V in the
triangular case and 2.05V in the square case. 
The difference between the on board measurements and the SPICE simulations
may be due to less than perfect current sinks. 
\begin{table}[h]
\caption{\baselineskip14pt \textbf{Voltage drop measurements}}
\smallskip\noindent
\centering
\begin{tabular}{ccc}
\toprule
&Square arrangement     &  Triangular arrangement      \\ 
\midrule
On board measurement & $2.03$V      &   $1.91$V       \\
\midrule
SPICE simulation&$2.05$V       &   $1.94$V         \\
\bottomrule
\end{tabular}
\label{Table1}
\end{table}
%
%
\section{Analytic expression for the voltage drop in the square and
triangular pad arrays}\label{MY}
In this section, analytic expressions for the voltage drop in both the square
and the triangular pad arrangements are obtained. 
Both configurations correspond to lattices in the plane, permitting direct use of the
standard theory of elliptic functions. We next set out those aspects of the
theory that we will need, as well as the special results that pertain for the
square and equilateral lattices, drawing on the classic text by Hille
\cite[Section~13.2]{Hille} as a standard general reference.

\subsection{The square and equilateral lattices}
A lattice of points in the complex plane consists of all integer linear
combinations $2w_1m + 2 w_3n$ ($m$, $n \in \Z$) of two given complex numbers
$2w_1$ and $2w_3$ for which $w_3/w_1$ has positive imaginary part. We
immediately specialize to the case in which 
\beq\label{4.1}
2w_1 = d >0 \ \mbox{ and }\ 2w_3 = \alpha d \mbox{ where }\alpha  = 
e^{2\pi i/q},\ q\in \mathbb{N},
\eeq
so that $\alpha$ is a $q^{\rm th}$-root of unity.
In this case the lattice is described by
\beq
\Lambda = \big\{ \lambda_{m,n} = md + n\alpha d:\ m,\, n \in \Z\big\}.
\eeq
The resulting lattice is invariant under multiplication by $\alpha$ 
precisely when there are integers $k$ and $j$ such that
\[
e^{4\pi i/q} = \alpha^2 = k\alpha + j = k e^{2\pi i/q} + j.
\]
It is not difficult to see, for example by examining the resulting equations for
the real and imaginary parts separately, that $\alpha$ will satisfy such an
identity only in the cases $q=4$ and $q=6$. The case $q=4$, with $\alpha = i$,
$2w_3 = id$, $\alpha^2 = -1$, corresponds to the square lattice. The case $q=6$,
with $\alpha = e^{\pi i/3}$, $2w_3 = e^{\pi i/3}d$,  $\alpha^2 = \alpha-1$,
corresponds to the triangular lattice. The values of $d$ in each case are
governed by \eqref{2.2} which guarantee that the areal densities of the pads
agree. 

Much of the analysis in the next subsections is essentially unchanged whether we 
work with the square or with the triangular lattice. 
We will therefore retain the notation $q$, $d$, $\alpha$ with the understanding that 
\beq\label{4.2}
(q,d,\alpha) = 
	\left\{
	\begin{array}{ll}
		(4,1,i)	& \mbox{ in the case of the square lattice,}  \\
		\displaystyle \Big(6,\frac{\sqrt{2}}{\root 4\of 3},e^{\pi i/3}\Big) & 
					\mbox{ in the case of the triangular lattice, }\\
  	\end{array}
	\right.
\eeq
the advantage being that in this way we can treat both configurations simultaneously.

\subsection{The Weierstrass $\sigma$-function for a plane lattice}
The Weierstrass elliptic $\mathcal{P}$-function associated with a lattice 
is doubly periodic with periods $2w_1$ and $2w_3$, 
and is analytic except for double poles at each of the lattice points. 
The corresponding $\sigma$-function is defined by 
\beq\label{sigmafn}
\sigma(z) = z \prodprime\left( 1-\frac{z}{\lambda} \right) 
\exp\left( \frac{z}{\lambda} + \frac{z^2}{2\lambda^2} \right),
\eeq 
where $\prod\nolimits{'}$ denotes the product over all lattice points with zero omitted.
The Weierstrass~$\zeta$-function is defined by 
\beq\label{zetafn}
\zeta(z) = \frac{1}{z} + \sumprime \left( \frac{1}{z-\lambda} \ +\  \frac{1}{\lambda}\ +\ 
\frac{z}{\lambda^2}\right),
\eeq
where $\sum\nolimits{'}$ denotes the sum over all lattice points with 0 omitted.

A quasi-periodicity property of the function $\sigma$ plays a key role in our analysis. 
Set 
\[
\eta_1 = \zeta(w_1)\quad \mbox{ and }\quad \eta_3 = \zeta(w_3).
\]
Then, \cite[Identity 13.2.19]{Hille},
\[
\sigma( z+ 2w_k) = -e^{2\eta_k(z+w_k)}\,\sigma(z), 
\quad z\in \C, \ k=1,\,3.
\]
These two identities lead to the full quasi-periodicity property 
\beq\label{sigmaperiodicity1}
\sigma(z+ 2mw_1+2nw_3) = 
(-1)^{m+n+mn}\,\exp\big[ (z+mw_1+nw_3)(2m\eta_1+2n\eta_3)\big]\, \sigma(z),
\eeq
for any integers $m$ and $n$. 
To proceed further, we need to compute  $\eta_1$ and $\eta_3$ explicitly 
for the square and the triangular lattices, 
at which point the quasi-periodicity property 
\eqref{sigmaperiodicity1} will become explicit in these cases. 
While these results are 
known, we give the explicit computations here for completeness.

\subsection{Computation of $\eta_1$ and $\eta_3$ for the square and triangular lattices}
In the case of a general lattice, $\eta_1$ and $\eta_3$ are related by the identity 
$2w_3 \eta_1 - 2w_1 \eta_3 = i\pi$ \cite[Exercise ?]{Hille}.
In the case of either of our lattices, this identity becomes (see \eqref{4.1})
\beq\label{etaidentity1}
d\alpha \eta_1 - d \eta_3 = i\pi. 
\eeq
The invariance of the lattice under multiplication by $\alpha = e^{2\pi i/q}$ 
and its powers, with $q=4$ for the square lattice and $q=6$ for the triangular lattice, 
leads to a second linear relationship between $\eta_1$ and $\eta_3$ as follows. 
By definition,
\[
\eta_1 = \zeta\left( \frac{d}{2} \right) = 
	\frac{2}{d} + \sumprime \left( \frac{1}{d/2-\lambda} \ +\  \frac{1}{\lambda}\ +\ 
	\frac{d^2}{4\lambda^2}\right).
\]
Replacing $\lambda$ by $\alpha^k \lambda$, $k=1, \ldots,  q-1$, 
gives a total of $q$ expressions for $\eta_1$. 
Adding these leads to 
\[
\eta_1 = \frac{2}{d} + \frac{1}{q}\sumprime \ \sum_{k=0}^{q-1} 	
	\left( \frac{1}{d/2- \alpha^k\lambda} \ +\  \frac{1}{\alpha^k\lambda}\ +\ 
	\frac{d^2}{4\alpha^{2k}\lambda^2}\right).
\]
Since 
\beq\label{alpha0}
\sum_{k=0}^{q-1}\alpha^{-k} = 0  = \sum_{k=0}^{q-1}\alpha^{-2k},
\eeq
we find that 
\beq\label{eta1eq}
\eta_1 = \frac{2}{d} + S, \quad \mbox{where } S = \frac{1}{q}\sumprime \ \sum_{k=0}^{q-1} 
\frac{1}{d/2- \alpha^k\lambda}.
\eeq
This procedure is repeated for $\eta_3$, which is given by  
\[
\eta_3 = \zeta\left( \frac{d\alpha}{2} \right) = \frac{2}{d\alpha} + 
\sumprime \left( \frac{1}{d\alpha /2-\lambda} \ +\  \frac{1}{\lambda}\ +\ \frac{d^2\alpha^2}{4\lambda^2}\right).
\]
Replacing $\lambda$ by $\alpha^k \lambda$, $k=1, \ldots,  q-1$, and adding 
all $q$ expressions for $\eta_3$, leads to 
\begin{align}
\eta_3 & = \frac{2}{d\alpha} + \frac{1}{q}\sumprime \ \sum_{k=0}^{q-1} 
	\left( \frac{1}{d\alpha/2- \alpha^k\lambda} 
	\ +\  \frac{1}{\alpha^k\lambda}\ +\ \frac{d^2\alpha^2}{4\alpha^{2k}\lambda^2}\right).\nonumber \\
& = \frac{2}{d\alpha} + \frac{1}{q}\sumprime \ \sum_{k=0}^{q-1} \frac{1}{d\alpha/2- \alpha^k\lambda} \nonumber \\
& = \frac{2}{d\alpha} + \frac{1}{q\alpha}\sumprime \ \sum_{k=0}^{q-1} \frac{1}{d/2- \alpha^k\lambda} \nonumber \\
& = \frac{2}{d\alpha} + \frac{1}{\alpha}\,S
= \left(\frac{2}{d}+S\right)\, \frac{1}{\alpha}.\label{eta3eq}
\end{align}
Together, \eqref{eta1eq} and \eqref{eta3eq} yield 
\beq\label{etaidentity2}
\eta_1 = \alpha\,\eta_3.
\eeq
Solving the simultaneous equations \eqref{etaidentity1} and \eqref{etaidentity2} gives
\beq\label{etavalues}
\eta_1 = \frac{i\alpha\pi}{d(\alpha^2-1)} \quad \mbox{and}\quad \eta_3 = \frac{i \pi}{d(\alpha^2-1)}.
\eeq
\begin{lemma}\label{lemma1}
In the case of the square lattice ($d=1$, $\alpha = i$) 
\beq\label{etasquare}
\eta_1 = \frac{\pi}{2}\quad \mbox{and}\quad \eta_3 = -\frac{i\pi}{2},
\eeq
while in the case of the triangular lattice ($d=\sqrt{2}/\root 4 \of 3$, $\alpha = e^{\pi i/3}$),
\beq\label{etatriangle}
\eta_1 = \frac{\,\pi}{\sqrt{2}\root 4 \of {3}}\quad \mbox{and}\quad 
\eta_3 = \frac{\,\pi}{\sqrt{2}\root 4 \of {3}}\,e^{-\pi i/3}.
\eeq
\end{lemma}
These results are easily verified, in view of \eqref{4.1}, by replacing 
$\alpha$ by $i$ and $d$ by 1 in \eqref{etavalues} in the case of 
the square lattice to obtain \eqref{etasquare}. 
In the case of the triangular lattice, replace 
$\alpha$ by $e^{\pi i/3}$ in \eqref{etavalues} and use
\[
\alpha^2 -1 = -\frac{3}{2} + \frac{\sqrt{3}}{2}i 
	= \sqrt{3}\, i\left( \frac{1}{2} + \frac{\sqrt{3}}{2}i\right)
 	= \sqrt{3}\,i\alpha
\]
to obtain $\eta_1 = \pi / (\sqrt{3}d)$ and $\eta_3 = \pi / (\sqrt{3}d\alpha) = \pi \overline{\alpha}/ (\sqrt{3}d)$. 
Finally, set $d = \sqrt{2} / \root 4 \of 3$ to obtain \eqref{etatriangle}.

\subsection{Quasi-periodicity and true periodicity for the square and triangular
lattices}
These values for $\eta_1$ and $\eta_3$ lead to a simple form of 
the general quasi-periodicity relation \eqref{sigmaperiodicity1} for the $\sigma$-function 
in the case of the square and the triangular lattices.
Surprisingly, perhaps, this relation has the same form in both cases, thereby unifying the 
analysis required to derive an analytic expression for the IR-drop. 
With an eye to \eqref{sigmaperiodicity1}, recall that a general lattice point is 
$\lambda_{m,n} = 2w_1m + 2 w_3n = md + n\alpha d$, where $m$ and $n$ are integers. 
Then, in the case of the square lattice with $d=1$ and 
using the $\eta$-values given by \eqref{etasquare},
\[
\lambda_{m,n} = m +in \quad \mbox{ and } \quad
2m\eta_1+2n\eta_3 = m\pi - in\pi  =\pi\, \overline{\lambda_{m,n} }.
\]
In the case of the triangular lattice with $d=\sqrt{2}/\root 4 \of 3$, 
we use the $\eta$-values given by \eqref{etatriangle} to obtain
\begin{align*}
&\lambda_{m,n}  = \frac{\sqrt{2}}{\root 4 \of 3} \,m + \frac{\sqrt{2}}{\root 4 \of 3} \,e^{\pi i/3} n, \\
&2m\eta_1+2n\eta_3  = \frac{2m\pi}{\sqrt{2}\root 4 \of {3}} + 
\frac{2n\pi}{\sqrt{2}\root 4 \of {3}}e^{-\pi i/3}
= \pi \left( \frac{\sqrt{2}}{\root 4 \of 3}\, m + \frac{\sqrt{2}}{\root 4 \of 3}\,  e^{-\pi i/3} n \right) 
=\pi\, \overline{\lambda_{m,n} }.
\end{align*}
The quasi-periodicity property \eqref{sigmaperiodicity1} of the $\sigma$-function, 
in the case of either the square or the triangular lattice, therefore becomes
\begin{align}
\sigma\big(z+\lambda_{m,n}\big) & = (-1)^{m+n+mn} 
	\exp\big[ (z+\tfrac{1}{2}\lambda_{m,n})\,\pi\,\overline{\lambda_{m,n}}\, \big]\, \sigma(z)\nonumber\\
& = (-1)^{m+n+mn} 
	\exp\left[ \pi  \overline{\lambda_{m,n}}\,z  + \frac{\pi}{2}\, \vert \lambda_{m,n}\vert^2 
	 \right]\, \sigma(z).
\label{sigmaperiodicity2} 
\end{align}
This quasi-periodicity property of the $\sigma$-function leads to true periodicity of a related function.

\begin{lemma}\label{lemma2}
Set 
\beq\label{hdefn}
h(z) = -\frac{1}{2\pi} \log \vert \sigma(z)\vert  + \frac{1}{4} \,\vert z \vert^2, \qquad z \in \C \setminus \Lambda.
\eeq
In the case when either $\Lambda$ is the square lattice or the triangular lattice, 
$h$ is periodic in the sense that 
$h(z+\lambda) = h(z)$, for $z\in \C \setminus \Lambda$, $\lambda \in \Lambda$.

Furthermore, the value of $h$ doesn't change under reflection in any side 
of the relevant lattice. 
\end{lemma}
\begin{rmk}
The periodiciy of $h$ in the case of square or triangular lattices also follows 
from the results in \cite[Proposition 3.4]{GL} which builds upon work in
\cite{Hayman}.  
Gr\"ochenig and Lyubarskii have a more general periodicity result which is valid for
all lattices and involves an explicit normalization factor in terms of $\eta_1$
and $\eta_3$. The computation of $\eta_1$ and $\eta_3$
above shows that no normalization factor arises for triangular or square lattices.
\end{rmk}
\begin{proof}
Taking the logarithm of \eqref{sigmaperiodicity2} with $\lambda = \lambda_{m,n} \in \Lambda$ leads to 
\[
\log \big\vert \sigma(z+\lambda) \big\vert  
= \log \vert \sigma(z)\vert + 
		\mbox{Re} \left[  \pi \overline{\lambda} z + \frac{\pi}{2} \vert \lambda\vert^2  \right]
= \log \vert \sigma(z)\vert + 
		\frac{\pi}{2}  \left[  \vert \lambda\vert^2 + 2\, \mbox{Re} \big( \overline{\lambda}\, z \big) \right].
\]
But,
\[
\vert  \lambda \vert^2 + 2\, \mbox{Re} \big( \overline{\lambda}\, z \big) 
 	=  \vert z + \lambda \vert^2 - \vert z \vert^2,
\]
so that 
\[
\log \big\vert \sigma(z+\lambda) \big\vert  
	= \log \vert \sigma(z)\vert + \frac{\pi}{2}  \left[ \vert z + \lambda \vert^2 - \vert z \vert^2 \right]. 
	\qedhere
\]
This establishes the periodicity of $h$. 

To see that $h$ is invariant under reflection in any side of the lattice, it suffices to show that 
\beq\label{hsym}
h\big(\alpha^{2k}\,\overline{z}\big) = h(z)
\eeq
where $\alpha = i$ and $k = 0$ or $1$ in the case of the square lattice, 
while $\alpha = e^{\pi i /3}$ and  $k=0$, $1$ or $2$ in the case of the triangular lattice. 
In any of these cases, 
\[
\sigma \big(\alpha^{2k}\,\overline{z}\big)  = 
\alpha^{2k}\,\overline{z} \prodprime\left( 1-\frac{ \alpha^{2k}\,\overline{z} }{\lambda} \right) 
\exp\left( \frac{ \alpha^{2k}\,\overline{z} }{\lambda} + 
\frac{ \alpha^{4k}\,\overline{z}^ 2}{2\lambda^2} \right).
\]
The invariance of the lattice under multiplication by $\alpha^{2k}$, 
that is $\alpha^{2k}\, \Lambda = \Lambda$, and then its invariance 
under complex conjugation, shows that 
\begin{align*}
\sigma \big(\alpha^{2k}\,\overline{z}\big) & = 
\alpha^{2k}\,\overline{z} \prodprime\left( 1-\frac{ \overline{z} }{\lambda} \right) 
\exp\left( \frac{ \overline{z} }{\lambda} + 
\frac{ \overline{z}^ 2}{2\lambda^2} \right)\\
& = \alpha^{2k} \,\overline{\left(
z \prodprime\left( 1-\frac{ z }{\overline{\lambda}} \right) 
\exp\left( \frac{ z }{\overline{\lambda}} + 
\frac{ z^ 2}{2\overline{\lambda}^2} \right)\right)}\\
& = \alpha^{2k} \, \overline{\left(
z \prodprime\left( 1-\frac{ z }{\lambda} \right) 
\exp\left( \frac{ z }{\lambda} + 
\frac{ z^ 2}{2\lambda^2} \right)\right)}\\
& = \alpha^{2k}\,\overline{\sigma(z)}.
\end{align*}
On taking logarithms, the identity \eqref{hsym} follows. 
\end{proof}

\subsection{Analytic expressions for the IR-drop in the square and the
triangular arrangements}
 
Let $\Omega_\varepsilon = \C\setminus \bigcup_{\lambda \in \Lambda}
\overline{D}(\lambda,\varepsilon)$ denote the region formed by removing from the plane 
a closed disk of radius $\varepsilon$ about each lattice point. 
Our main result  gives an analytic bound for the 
voltage drop in both the square and triangular arrangement of the pads. 
It continues to be possible to analyse both configurations simultaneously, which we do. 
After stating and proving the analytic bound, we derive the explicit 
numerical bounds \eqref{VM} and \eqref{VY} which prove, in particular, that 
the equilateral disposition outperforms the square arrangement.

Before stating the main analytical result, Theorem~\ref{thm1}, we need an estimate on 
the $\sigma$-function near the origin. 

\begin{lemma}\label{lemma3}
For $\vert z \vert^q  \leq \tfrac{3}{5}$, 
\beq\label{lema3}
\big\vert\, \log\vert \sigma(z) \vert - \log\vert z \vert \,\big\vert
\leq A_q\, \big(\vert z \vert^q + \vert z \vert^{2q}\big), 
\eeq
where $A_q= \frac{1}{q} \sumprime \frac{1}{\vert \lambda \vert^q}$ 
and where $q$ is 4 or 6  depending on whether we are working with the square or the triangular lattice. 
Correct to eight decimal places, 
\beq\label{A4value}
A_4 = \frac{1}{4} \sumprime \frac{1}{(m^2 + n^2)^2} = 1.50670300
\eeq
and 
\beq\label{A6value}
A_6 = \frac{\sqrt{3}}{16} \sumprime \frac{1}{(m^2 + n^2+mn)^3} 
	= 0.69020942.
\eeq
\end{lemma}

\begin{proof}Recall the expression \eqref{sigmafn} for the $\sigma$-function. 
By the symmetry of the lattice under multiplication by $\alpha^k$, we see that
\[
\sigma (z)  = 
z\, \prodprime\left( 1-\frac{ z }{\alpha^k\lambda} \right) 
\exp\left( \frac{z }{\alpha^k\lambda} + 
\frac{ z^ 2 }{2\alpha^{2k}\lambda^2} \right),
\quad k=0,\ 1,\ \ldots,\ q-1.
\]
When these $q$ expressions for $\sigma(z)$ are multiplied together, one obtains 
\beq\label{sigma^p}
\sigma^q(z) = z^q\, \prodprime\, \prod_{k=0}^{q-1}\left( 1-\frac{ z }{\alpha^k\lambda} \right)
= z^q\, \prodprime\, \left( 1-\frac{ z^q }{\lambda^q} \right),
\eeq
where \eqref{alpha0} leads to the elimination of the exponential terms, 
and the identity 
\[
1-w^q = (1-w)\,\left(1-\frac{w}{\alpha}\right)\,\ldots 
\left(1-\frac{w}{\alpha^{q-1}}\right), \quad q \in \mathbb{N},
\]
was used at the last step in \eqref{sigma^p}.
Taking the logarithm of \eqref{sigma^p} leads to 
\beq\label{logsigma}
\log\vert \sigma(z) \vert = \log\vert z \vert + 
\frac{1}{q} \sumprime \log \left\vert 1-\frac{ z^q }{\lambda^q} \right\vert .
\eeq
The power series expansion of the analytic function $-\log(1-w)$ about $0$ is 
\[
-\log(1-w) = w + \frac{w^2}{2} + \frac{w^3}{3} + \frac{w^4}{4} + \cdots, 
\]
so that, for $\vert w \vert  \leq \frac{3}{5}$, 
\begin{align}
\big\vert \log \vert 1-w\vert\hskip1pt \big\vert & = \big\vert {\rm Re}\big( \log (1-w) \,\big)\big\vert\nonumber\\
& \leq \big\vert \log (1-w) \big\vert\nonumber \\
& \leq \vert w\vert  + \frac{\,\vert w\vert^2}{2} + \frac{\,\vert w\vert ^3}{3} + 
	\frac{\,\vert w\vert ^4}{4} + \cdots\nonumber \\
&\leq \vert w \vert \left( 1 + \frac{\vert w \vert}{2} + \frac{\vert w\vert^2}{3}\,\frac{1}{1-\vert w\vert} 
	\right) \ \leq\ (1+\vert w \vert)\,\vert w \vert.\label{zz}
\end{align}
Since $\vert \lambda \vert \geq 1$ for $\lambda \in \Lambda \setminus \{0\}$, 
once $\vert z \vert^q  \leq \frac{3}{5}$ we can apply \eqref{zz} with $w  = (z/\lambda)^q$ to obtain
\[
\left\vert \frac{1}{q} \sumprime 
	\log \left\vert 1-\frac{ z^q }{\lambda^q} \right\vert \right\vert
\leq  \frac{1}{q} \sumprime \left( 
	\frac{\vert z \vert ^q}{\vert \lambda \vert^q} + 
	\frac{\vert z \vert ^{2q}}{\vert \lambda \vert^{2q}} \right)
\leq  A_q\, \big(\vert z \vert^q + \vert z \vert^{2q}\big), 
\]
where
\[
A_q= \frac{1}{q} \sumprime \frac{1}{\vert \lambda \vert^q}.
\]
Together with \eqref{logsigma}, this proves \eqref{lema3}. The estimates \eqref{A4value} and 
\eqref{A6value} can be obtained numerically. 
\end{proof}

\begin{thm}\label{thm1}
In the case of either the square or the triangular lattice, in each case with the values given in \eqref{4.2},
the solution of 
\beq\label{pde}
\left\{ \begin{aligned}
\Delta u_\varepsilon &\ =\ 1 \quad \rm{ in }\ \Omega_\varepsilon\\
u_\varepsilon &\ =\ 0 \quad \rm{ on }\ \partial \Omega_\varepsilon
\end{aligned}
\right.
\eeq
may be written as 
\beq\label{uepsilon}
u_\varepsilon(z) \ =\ -\frac{1}{2\pi}\log \vert \sigma(z)\vert + \frac{1}{4}\vert z \vert^2 
	 + \frac{1}{2\pi} \log\varepsilon - \frac{1}{4}\varepsilon^2  + h_{\varepsilon}(z),
\eeq
where $h_\varepsilon$ satisfies
\beq\label{hbound}
\big\vert h_\varepsilon(z)\big\vert \ \leq\ \frac{A_q}{2\pi} \, \big( \varepsilon^q
	+ \varepsilon^{2q} \big), \qquad z\in \Omega_\varepsilon,
\eeq
and $A_q$ has the value given in the statement of Lemma~\ref{lemma3}.
\end{thm}
\begin{proof}
Let $h_\varepsilon$ be the function which is harmonic on $\Omega_\varepsilon$ 
and has boundary values
\beq\label{hvalues}
h_\varepsilon(\zeta) = \frac{1}{2\pi}\log \vert \sigma(\zeta)\vert -
	 \frac{1}{4}\vert \zeta \vert^2  -  \frac{1}{2\pi} \log\varepsilon + \frac{1}{4}\varepsilon^2,
	 \quad \zeta\in \partial \Omega_\varepsilon.
\eeq
By Lemma~\ref{lemma2}, these boundary values are periodic and therefore so too 
is $h_\varepsilon$ (that is, $h_\varepsilon(z+\lambda) = h_\varepsilon(z)$
for $z\in \Omega_\varepsilon$ and $\lambda \in \Lambda$). 

Define a function $u_\varepsilon$ by \eqref{uepsilon}. 
Then $\Delta u_\varepsilon = 1$ in $\Omega_\varepsilon$, this because
$\Delta \big( \vert z \vert^2 \big) = 4$ while 
$\log \vert \sigma(z)\vert$ is harmonic on $\Omega_\varepsilon$ 
being the logarithm of the modulus of a non-vanishing analytic function there. 
Moreover, $u_\varepsilon$ vanishes on the boundary of $\Omega_\varepsilon$, 
so that $u_\varepsilon$ is the solution of \eqref{pde}. 

Set $D_0$ to be the interior of the square with vertices $0$, $1$, $1+i$ and $i$ in the case 
of the square lattice and set $D_0$ to be the interior of the triangle with vertices $0$, 
$\sqrt{2}/\root 4\of 3$ and $\sqrt{2}e^{\pi i/3}/\root 4\of 3$ in the case of the triangular lattice.
The bound \eqref{hbound} for $h_\varepsilon$ is obtained by applying
the maximum principle to $h_\varepsilon$ on $\Omega_\varepsilon \cap D_0$. 
If $h_\varepsilon$ were to assume an extremal value on the closure of $\Omega_\varepsilon \cap D_0$ 
at a point of $\Omega_\varepsilon \cap \partial D_0$ then, by the symmetry of $h_\varepsilon$ 
in the sides of $D_0$ (see the final part of Lemma~\ref{lemma2}),
$h_\varepsilon$ would have a local extremum there, contradicting the maximum principle. 
Thus $h_\varepsilon$ achieves its extremum values (over $\Omega_\varepsilon$ or,
equivalently, over $\Omega_\varepsilon \cap D_0$) at a point of 
$\partial \Omega_\varepsilon$ that is, again using the periodicity of $h_\varepsilon$,
at a point of $C(0,\varepsilon)$. 
Taking account of the boundary values \eqref{hvalues} and then Lemma~\ref{lemma3} 
we see that, for $\vert \zeta \vert = \varepsilon$, 
\[
\big\vert h_\varepsilon(\zeta) \big\vert = \frac{1}{2\pi}\big\vert 
	\log \vert \sigma(\zeta)\vert  -   \log\varepsilon
	\big\vert 
\leq \frac{A_q}{2\pi}\,\big(\varepsilon^p + \varepsilon^{2p}\big) .
\]
Thus, by the maximum principle, the harmonic function $h_\varepsilon$ 
satisfies the bound \eqref{hbound} throughout $\Omega_\varepsilon$. 
\end{proof}

\begin{thm}\label{thm2}
The maximum voltage drop $\VM(\varepsilon)$, when the pads
are arranged in a square lattice and with the parameters given in \eqref{4.2},
satisfies 
\beq\label{Mdrop2}
\left\vert V_{\max}^M(\varepsilon) 
	- \bigg[\frac{1}{2\pi}\log \frac{1}{\varepsilon} - C_M + \frac{1}{4}\varepsilon^2\bigg] 
	\right\vert 
	\ \leq\ \frac{A_4}{2\pi} \big( \varepsilon^4 + \varepsilon^8 \big),
\eeq
where $A_4$ is given by \eqref{A4value} and
\beq\label{CM}
C_M\ = \ \frac{1}{\pi}\log \Gamma(\tfrac{1}{4}) - \frac{1}{2\pi}\log(2\sqrt{2\pi}) 
\ =\  0.1534 1889 3205,
\eeq
correct to 12 decimal places.

\noindent The maximum voltage drop $\VY(\varepsilon)$, when the pads
are arranged in an equilateral triangular lattice and with the parameters given in \eqref{4.2},
satisfies 
\beq\label{Ydrop2}
\left\vert V_{\max}^Y(\varepsilon) 
	- \bigg[\frac{1}{2\pi}\log \frac{1}{\varepsilon} - C_Y + \frac{1}{4}\varepsilon^2\bigg] 
	\right\vert 
	\ \leq\ \frac{A_6}{2\pi} \big( \varepsilon^6 + \varepsilon^{12} \big),
\eeq
where $A_6$ is given by \eqref{A6value} and
\beq\label{CY}
C_Y\ = \ 
	\frac{3}{2\pi}\log \Gamma(\tfrac{1}{3}) - \frac{1}{2\pi}\log(2\sqrt{2}\pi) +\frac{1}{8\pi} \log 3  
\ =\ 0.166549975068,
\eeq
correct to 12 decimal places.
\end{thm}
\begin{proof}
In the case of the square arrangement of pads, 
the maximum voltage drop occurs at the point $b_s=(1+i)/2$ which 
lies at the centre of the square formed by the lattice points at $0$, $1$, $1+i$ and $i$ 
(see Section~\ref{sec6}). 
The negative of the expression \eqref{uepsilon}, evaluated at $z=b_s$, is the maximum 
voltage drop. Since $\vert b_s \vert^2 = 1/2$,
\[
\VM \ =\ -u_\varepsilon(b_s) \ =\  
	\frac{1}{2\pi}\log \frac{1}{\varepsilon} - C_M + \frac{1}{4}\varepsilon^2 -h_\varepsilon(b_s)
\]
where
\beq\label{CM2}
C_M = \frac{1}{8} - \frac{1}{2\pi}\log \left\vert \sigma\left( b_s\right) \right\vert.
\eeq
Formulas 18.14.7 and 18.14.9 in Abramowitz and Stegun \cite{AbramStegun} give
\[
\sigma(w_2) = \sqrt{2}\,e^{(1+i)\pi/4} \quad \mbox{when} \quad 
w_1 = \frac{\Gamma^2(\frac{1}{4})}{4\sqrt{\pi}},\ w_3 = iw_1, \ w_2 = w_1+w_3.
\]
Scaling by $t = 2\sqrt{\pi} / \Gamma^2(\tfrac{1}{4})$ so that $w_1 = 1/2$, 
and noting that the function $\sigma$ also scales linearly, we find that 
\[
\sigma(b_s) = \frac{2\sqrt{\pi}}{\Gamma^2(\frac{1}{4})}\, \,\sqrt{2}\,e^{(1+i)\pi/4}
\]
Then,
\[
\log \vert \sigma(b_s)\vert = \frac{\pi}{4} + \log(2\sqrt{2\pi}) -2 \log \Gamma(\tfrac{1}{4}),
\]
so that \eqref{CM} follows from \eqref{CM2}, and then \eqref{Mdrop2} follows from 
the bound \eqref{hbound} for $h_s$.
 
In the case of the triangular pad arrangement, the maximum voltage drop 
occurs at the point $b_t = 3^{-3/4}\sqrt{2} e^{\pi i/6}$ 
which lies at the centre of the equilateral triangle with vertices $0$, $2w_1 = d$,  
$2w_3 = d\,\alpha$, where $d = \sqrt{2}/\root 4 \of 3$ and 
$\alpha = e^{\pi i/3}$.
Since $\vert b_t \vert^2 = 2/(3\sqrt{3})$, 
\[
\VY(\varepsilon) \ =\ -u_\varepsilon(b_t) \ =\  
	\frac{1}{2\pi}\log \frac{1}{\varepsilon} - C_Y + \frac{1}{4}\varepsilon^2 -h_\varepsilon(b_t)
\]
where
\beq\label{CY2}
C_Y = \frac{1}{6\sqrt{3}} -  \frac{1}{2\pi}\log \left\vert \sigma\left( b_t\right) \right\vert.
\eeq
Formulas 18.13.15 and 18.13.28 in Abramowitz and Stegun \cite{AbramStegun} give
the value of $\sigma$ at the centre of the equilateral triangle as
\[
e^{\pi/(3\sqrt{3})}\,e^{i\pi/6} \quad \mbox{when} \quad 
	w_1 = \frac{\Gamma^3(\frac{1}{3})}{4\pi} 
	\mbox{ and }\ w_3 = iw_1.
\]
Scaling by $t = 2\pi d / \Gamma^3(\tfrac{1}{3})$ leads to 
\[
\sigma(b_t) = \frac{2\sqrt{2}\pi}{\root 4 \of 3\, \Gamma^3(\tfrac{1}{3})} \,
e^{\pi/(3\sqrt{3})}\,e^{i\pi/6}.
\]
Then,
\[
\log \vert \sigma(b_t)\vert = \frac{\pi}{3\sqrt{3}} + \log(2\sqrt{2}\pi) - \frac{1}{4}\log 3 -3 \log \Gamma(\tfrac{1}{3}),
\]
\eqref{CY} follows from \eqref{CY2}, and then \eqref{Ydrop2} again follows from 
the bound \eqref{hbound} for $h_s$.
\end{proof}
%
%
\section{The Hexagonal configuration}\label{H}
We  estimate the voltage drop for the hexagonal lattice  
with the same aerial density of pads as in the case of the 
square and the equilateral power pad arrangements 
analysed in the previous section.  
The geometric setting is the following. We consider the 
domain 
\[
\Omega=\Omega_\varepsilon=\C\setminus \bigcup_{\lambda\in\Lambda}
D(\lambda,\varepsilon)
\]
where $\Lambda$ is the set of vertices of the blue hexagonal grid shown in Figure~\ref{fig2}.

It will be convenient to consider the set of centres $\Lambda$ as the
difference of two lattices: see Figure~\ref{fig2}. 
The first lattice consists of the black \textsl{and\/} the red vertices in Figure~\ref{fig2}, 
which we denote by $BR$, 
while the second lattice consists of the red vertices alone, which we denote by $R$.
Thus $\Lambda=BR\setminus R$. 
Both $BR$ and $R$ are lattices that determine an equilateral grid. 
The main advantage of considering $\Lambda$ as a difference
of two lattices is that for any equilateral lattice we can construct an associated
Weierstrass entire function with zeros on the lattice whose pseudo-periodicity properties 
were analysed in the previous section.  
Thus, instead of directly building an entire function with zeros on $\Lambda$ 
we obtain more information by considering a
quotient of two entire functions, one vanishing on $BR$ and the other on $R$. 

The maximum voltage drop corresponds to the minimum value of $u$,
where $u$ is the solution to $\Delta u =1$ in $\Omega_\varepsilon$ and $u=0$ in
the boundary of $\Omega$. 
The maximum voltage drop is, consequently, at least as big as $-u(0)$ where $0$
is 
at the centre of a hexagon. 

Let us denote by $\sigma(z)$ the Weierstrass $\sigma$-function associated with 
the equilateral triangular lattice with side length 
$d_2 = \sqrt{2}/\root 4 \of 3$ as described in  \eqref{4.2}. 
The $\sigma$-function for the lattice $BR$, 
with sidelength $d_3 = 2/\root 4 \of {27}$, is then 
\[
\sigma_{BR}(z)\ =\ \frac{d_3}{d_2}\,\sigma\bigg( \frac{d_2}{d_3} \,z\bigg)
\ =\ \sqrt{\tfrac{2}{3}}\,\sigma\bigg( \sqrt{\tfrac{3}{2}} \,z\bigg)
\] 
while the $\sigma$-function for the lattice $R$,  with sidelength $\sqrt{3}d_3$, 
is  
\[
\sigma_R(z) \ =\ \sqrt{2}\, \beta\,\sigma\bigg( \frac{1}{\sqrt{2}\,\beta} \,z\bigg),
\mbox{ where } \beta=e^{\pi i/6}.
\]
Clearly $\sigma_R$ vanishes on the vertices of $R$ and $\sigma_{BR}$ 
vanishes on the vertices of $BR$.

Consider the function defined in $\Omega$ by
\[
v(z)= v_{BR}(z)-v_R(z) = \left[ \frac 3{8} |z|^2-\frac 1{2\pi} \log|\sigma_{BR}(z)| 
	- c_\varepsilon\right] -
	\left[\frac 1{8} |z|^2-\frac 1{2\pi}\log|\sigma_R(z)|  -d_\varepsilon \right]
\]
where $c_\varepsilon$ and $d_\varepsilon$ are to be chosen appropriately.
Both functions $v_{BR}$ and $v_R$ have many symmetries. In
particular they are symmetric across any line that extends any of the
sides of the hexagon which form the original grid. 
Thus $v$ has the same symmetry.
Moreover $\Delta v=1$ in $\Omega$, so that $v$ is close to the desired 
solution $u$ to the problem. 
In fact, they differ by a harmonic function, in that $u=v+h$. 
The desired value $u(0)$ can be approximated by the value of  $v$ 
at the centre of the hexagon. 
The error that we make, that is $h(0)$, can again be estimated by 
the maximum principle, in that
$|h(0)|\le \sup_{\partial \Omega} |h|= \sup_{\partial \Omega} |v|$. 

The constants $c_\varepsilon$ and $d_\varepsilon$ will now be chosen so that
both $\sup_{\partial \Omega}v_{BR}$ and $\sup_{\partial \Omega}v_R$ are small.
The selection of $c_\varepsilon$ required to make $v_{BR}$ 
small on the boundary is the more straightforward. 
By the symmetries of $v_{BR}$, 
$\sup_{\partial\Omega}v_{BR}=\sup_{\partial D(0,\varepsilon)}v_{BR}$. 
Observe that although $\partial D(0,\varepsilon)$ is not 
part of the boundary of $\Omega$,
all disks around the vertices of the combined black and red triangular grid 
are equal if we restrict our attention to $v_{BR}$. 
On $\partial D(0,\varepsilon)$ the value of $v_{BR}$ is close to a constant. 
In fact we see from Lemma~\ref{lemma3} that
\[
 v_{BR}(z)= \frac 3{8}\varepsilon^2+\frac 1{2\pi} \log \frac 1{\varepsilon} +
O(\varepsilon^6)  - c_\varepsilon,
\quad \mbox{for all }z\in D(0,\varepsilon).
\]
Thus, with the choice of 
$c_\varepsilon=\frac 1{2\pi} \log \frac 1\varepsilon +\frac 3{8}
\varepsilon^2$,
we obtain that $\vert v_{BR}(z)\vert \le C \varepsilon ^6$ on 
 $\partial \Omega_\varepsilon$.

We consider now the values of $v_R$ on the boundary of 
$\Omega_\varepsilon$ which consists
of disks of radius $\varepsilon$ centred at the baricentres of the red triangles. 
The function $v_R$ has the same behaviour at each. 
Let us denote one of the baricentres by $A$. 
Then, as established in Section~\ref{sec6}, $v_R$ has a local minimun at $A$. 
We can actually prove that
\[
 v_R(z) = v_R(A) + \frac {1}{8} \vert z-A\vert^2 + O(\varepsilon^3), 
 \quad \mbox{for all }z\in D(A, \varepsilon).
\]
Thus, if we choose $d_\varepsilon= \frac18 |A|^2 -\frac 1{2\pi} \log
|\sigma_R(A)| +\frac 18 \varepsilon^2$, 
then $|v_R(z)| \le C \varepsilon ^3$
on $\partial D(A,\varepsilon)$ and therefore on $\partial\Omega_\varepsilon$.

Finally we have proved that 
$\sup_{\partial\Omega_\varepsilon}
\vert h \vert =\sup_{\partial\Omega_\varepsilon}|v|\le C\varepsilon^3$.
The voltage drop at the centre of a hexagon is $-u(0)=-v(0)-h(0)$, and so 
\[
\VH(\varepsilon) := -u(0) = c_\varepsilon-d_\varepsilon + O(\varepsilon^3)	
	= \frac 1{2\pi} \log \frac1\varepsilon - \frac18 \vert A\vert^2 
	+\frac1{2\pi} \log \vert \sigma_R(A)\vert 
	+ \frac {1}{4} \varepsilon^2+ O(\varepsilon^3).
\]
Observe that 
$\log |\sigma_R(A)|= \log{\sqrt{2}}+ \log|\sigma(A/(\alpha\sqrt{2}))|$. 
In our setting $|A|^2=4/(3\sqrt3)$ and
the value of the $\sigma$ function on the baricenter of its defining triangle
can be computed explicitly, see Abramowitz-Stegun formula 18.13.28, as 
\[
\bigg\vert \sigma\bigg(\frac{A}{\beta\sqrt{2}} \bigg)\bigg\vert
	=e^{\pi/(3\sqrt{3})} \frac{2\sqrt{2}\pi}{3^{1/4}\Gamma(1/3)^3 }
	\simeq 0.642836690101
\]
Thus the voltage drop at the centre of a hexagon is 
\[
\VH(\varepsilon)  = \frac 1{2\pi} \log \frac1\varepsilon 
	-  0.111391075030 + \frac 14 \varepsilon^2 + O(\varepsilon^3),
\]
which is \eqref{VH}. 
The conclusion is that the hexagonal grid has the worst voltage drop
among the ones that we considered, with the best being the triangular lattice and
the standard square lattice being in an intermediate position.

\section{Where does the maximum voltage drop occur?}
\label{sec6}
We examine now where the maximal voltage drop takes place in the square lattice
configuration and in the equilateral setting. Heuristically, one expects the
voltage drop to be maximal in the center of the squares and in the barycenter
respectively. This has been taken for granted in the literature, but we will
nevertheless give a rigorous proof of this intuitive fact. The case of the
square is the easiest one.
\begin{proof}
 Consider the solution $v$ in the unbounded domain $\Omega$ to the mixed
Dirichlet-Neumann problem as in the figure~\ref{square}:
\begin{figure}
\caption{Auxiliary domain for the square lattice}\label{square}
\begin{center}
\begin{tikzpicture}[y=-1cm]

\draw[black] (-6.29824,-2.80444) +(-90:0.61334) arc (-90:0:0.61334);
\draw[black] (-6.29825,-6.5) +(0:0.61381) arc (0:90:0.61381);
\draw[black] (-2.59819,-6.50444) +(180:0.6196) arc (180:90:0.6196);
\draw[black] (-2.6032,-2.80444) +(180:0.61459) arc (180:270:0.61459);
\draw[black] (-5.68444,-2.8) -- (-3.21778,-2.8);
\draw[black] (-5.68444,-6.50222) -- (-3.21778,-6.50222);
\draw[black] (-13.70114,-2.80222) +(-90:0.61556) arc (-90:0:0.61556);
\draw[black] (-13.70404,-6.50444) +(0:0.6196) arc (0:90:0.6196);
\draw[black] (-10.00397,-6.5) +(-180:0.61381) arc (-180:-270:0.61381);
\draw[black] (-10.00321,-2.80444) +(180:0.61458) arc (180:270:0.61458);
\draw[black] (-13.08444,-2.8) -- (-10.61778,-2.8);
\draw[black] (-13.08444,-6.50222) -- (-10.61778,-6.50222);
\draw[black] (-6.30396,-2.80444) +(-90:0.61335) arc (-90:-180:0.61335);
\draw[black] (-6.30396,-6.5) +(-180:0.61382) arc (-180:-270:0.61382);
\draw[black] (-9.99826,-6.5) +(0:0.61382) arc (0:90:0.61382);
\draw[black] (-9.99902,-2.80444) +(0:0.61459) arc (0:-90:0.61459);
\draw[black] (-6.91778,-2.8) -- (-9.38444,-2.8);
\draw[black] (-6.91778,-6.50222) -- (-9.38444,-6.50222);
\path (-8.56222,-5.34) node[text=black,anchor=base west] 
{$\Delta v=1$};
\path (-8.37556,-7.40222) node[text=black,anchor=base west] 
{$v_y=0$};
\filldraw[black] (-8.15111,-4.65111) circle (0.06222cm);
\filldraw[black,arrows=-to] (-10.00222,-6.87111) -- (-9.57778,-6.23778);
\draw[dashed,black] (-2.6,-4.65111) -- (-13.70222,-4.65111);
\filldraw[black,arrows=-to] (-7.98889,-7.24444) -- (-7.98889,-6.54889);
\path (-8.32222,-3.66) node[text=black,anchor=base west] 
{$\Omega$};
\path (-10.3,-7) node[text=black,anchor=base west] 
{$v=0$};
\path (-8.0,-4.2) node[text=black,anchor=base west] 
{$0$};
\end{tikzpicture}%
\end{center}
\end{figure}
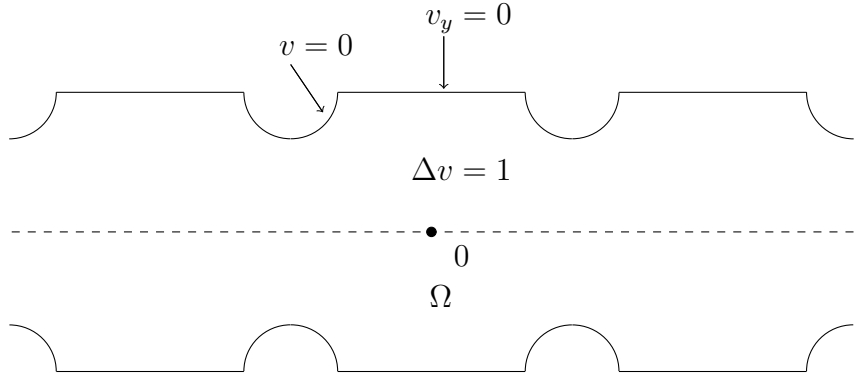
We want to prove that it has a minimum at $z=0$. We will prove that 
the function $v_y>0$ when $\Im z>0$ and $v_y<0$ when $\Im z <0$.

Clearly 
\[
\Delta v_y=\frac{\partial \Delta v}{\partial y}= \frac{\partial 1}{\partial
y}=0.
\]
Thus $v_y$ is harmonic. Moreover in the ``straight'' pieces of the boundary
$v_y=0$. On the half circles $v=0$, thus $\nabla v$ is perpendicular to the
circles. Therefore $v_y=\langle \nabla v, (1,0)\rangle$ is positive in the half
circles to the top of the dotted line and negative in the others. By symmetry
$v_y=0$ on the dotted line. Thus solving the Dirichlet problem for $v_y$ in the
domain $\Omega_+:=\Omega\cap \{\Im z>0\}$ we see that $v_y\le 0$ in $\Omega_+$
(in the boundary it is positive and $v_y\le 0$ in $\Omega_-:=\Omega\cap \{\Im
z<0\}$

We do similarly along the $x$-direction and we are done
\end{proof}

In the case of the triangular lattice we consider the domain as in 
figure~\ref{triangular}. The domain $\Omega$ is the equilateral triangle where
we remove the three disks of equal radius centered at the corners of the
triangle. Let $p$ be the barycenter of the triangle and define the function $u$
such that $\Delta u=1$ in the interior of $\Omega$, $u=0$ in the part of the
boundary of $\Omega$ defined by the arcs of circle and $\partial u /\partial
n=0$ in the part of the boundary of $\Omega$ defined by the sides of the
triangle. The claim is the following:
\begin{claim}
 There is only a minimum value of $u$ in $\Omega$ and it is attained at $p$.
\end{claim}
\begin{proof}
 We will make this argument by a variation on the radius of the disks. It will
be convenient to denote the domains $\Omega_t$ to the domain obtained removing
the disks of radius $t$ and $u^t$ the corresponding solution. We will denote by
$v$ the Green function of the flat torus whose fundamental domain is twice the
equilateral triangle. It follows from the  definition that the Green function of
this torus is the function $v(z)=\frac 14 |z|^2 -\frac{1}{2\pi} \log
|\sigma(z)|$ as we saw in Lemma~\ref{lemma2}. In a sense we will see that
$u_t$ is very close to $v$ as $t\to 0$. We are interested in the critical
points of $u^t$. The corresponding critical points for $v$ have been identified
in \cite{LinLung} and the only ones appearing are the trivial ones that can be
identified by symmetry considerations. There is a local minimum of $v$ at $p$
and three saddle points in the midpoints of the sides of the triangle. 

We are going to prove that a very similar structure arises in the case of
$u_t$: There is a minimum at $p$ and three saddle points at the midpoints of
the sides of the triangle. 

Along all this discussion we will restrict ourselves to the case $0<t<t_0$
where $t_0$ is the biggest radius such that the disks defining $\Omega_t$ are
disjoint since this is the only relevant case.

We start by observing that at the barycenter $p$ there is a critical point for
$u^t$ for symmetry reasons. Moreover since $u^t(e^{2\pi i /3}(z-p))=u_t(z-p)$ 
the
Hessian of $u$ at $p$ must be a constant times the identity matrix. Since
$\Delta u(p)=1$ it follows that $u_{xx}(p)=u_{yy}(p)=1/2$.

\begin{figure}
\caption{The equilateral fundamental domain}\label{triangular}
\begin{center}
\begin{tikzpicture}[y=-1cm]

\path[fill=yellow!55] (6.58889,5.03778) -- (6.58222,7.30667) -- 
(8.51333,8.40222) -- (6.59556,5.04444) -- cycle;

\draw[black] (6.58667,7.30444) -- (6.58667,5.02222);
\draw[black] (6.58667,7.30444) -- (4.61556,8.43778);
\draw[black] (6.58667,7.30444) -- (8.55556,8.43778);
\draw[black] (6.58667,5.02222) -- (4.60889,8.44444) -- (8.56222,8.44444) -- 
cycle;
\draw[dashed,black] (6.58667,7.30444) -- (7.56444,6.72);
\draw[dashed,black] (6.58667,7.30444) -- (5.6,6.73333);
\draw[dashed,black] (6.58667,7.30444) -- (6.58667,8.44222);
\path (6.2,7.7) node[text=black,anchor=base west] {$p$};
\path (7.3,7.69333) node[text=black,anchor=base west] {$d_1$};
\path (6.5,6.58) node[text=black,anchor=base west] {$d_2$};
\path (5.73556,8.01778) node[text=black,anchor=base west] {$d_3$};

\path[draw=black,fill=white] (6.57778,5.03111) circle (0.41556cm);
\path[draw=black,fill=white] (4.61111,8.44444) circle (0.41556cm);
\path[draw=black,fill=white] (8.55778,8.44444) circle (0.41556cm);

\path (4.3,8.56444) node[text=black,anchor=base west] {$v_3$};
\path (6.3,5.07556) node[text=black,anchor=base west] {$v_2$};
\path (8.3,8.50222) node[text=black,anchor=base west] {$v_1$};
\path (5.9,6.88222) node[text=black,anchor=base west] {$o_2$};
\path (6.5,7.92222) node[text=black,anchor=base west] {$o_3$};
\path (6.8,6.91333) node[text=black,anchor=base west] {$o_1$};

\end{tikzpicture}%
\end{center}
\end{figure}
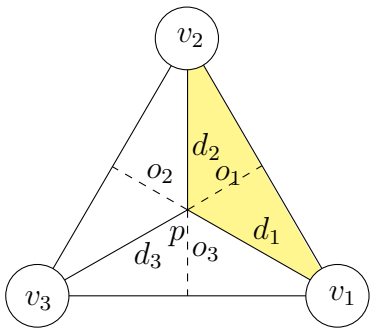

Let $d_1,d_2,d_3$ be vectors pointing from $p$ to one of the vertex of the
triangle as in Figure~\ref{triangular}. By symmetry again the gradient of $u^t$
in any point of the median of the triangle is a multiple of $d_j$. 

Assume, for the
moment being, that there is a $\delta$ such that  for a given $t<\delta$ we have
proven that $u^t_{d_j}(x)>0$ for any $x$ in the median joining $p$ with a vertex
(excluding the barycenter), i.e., along the median the gradient is pointing
towards the vertices. 

Under this assumption we concentrate our attention on
the yellow region in the picture consisting of one third of the original domain
$\Omega_t$ limited by two of the medians. We will prove that on the yellow
region the function $u_{o_1}$ which is the derivative of $u$ in the direction
$o_1:=-d_3$ if strictly positive. This is clear because the function $u_{o_1}^t$
is an harmonic function ($\Delta u_{o_1}^t=0$) and in the boundary of the 
shaded region it is positive: on the medians it is positive by assumption, on 
the sides
of the triangle it is actually 0 by the definition of $u^t$ and on the arcs of
circles the gradient of $u^t$ is pointing towards the center of the disks
($u^t\equiv 0$ on the boundary of the disks and it is negative in $\Omega_t$),
thus $u_{o_1}$ is positive on the arcs of circle that limit the shaded region. 

Now any point $q$ belonging to the yellow region has the property that
$u(p)<u(q)$ since we can follow a path from $p$ to $q$ consisting of segment
over the median followed by a segment in the direction of $o_1$ and in both
segments $u^t$ will be increasing. 

It remains to prove that $u^t_{d_j}(x)\ge 0$  on the corresponding median. Let
us assume for the moment being that this is the case for all $t\le \delta$. We
will prove then that this is true for all $t<t_0$. 

Let us denote by $t^*$ the biggest $t$ such that $u^t_{d_j}(x)\ge 0$ on all 
points of the
median. We will see now that if $t^*<t_0$  we reach a contradiction. By
continuity $u^{t^*}_{d_j}(x)\ge 0$ on the median. If we prove that actually 
\begin{equation}\label{strict}
u^{t^{*}}_{d_j}(x)>\delta>0  
\end{equation}
 on the median we would have  reached a contradiction since $t^*$ 
would not be maximal. We cannot prove \eqref{strict} directly since 
$u_{d_j}(p)=0$, but in a neighborhood of $p$
$u_{d_j}(x)>u_{d_j}(p)$ since $u_{d_jd_j}^{t}(p)=1/2$. Thus if $t^*$ is maximal
it maybe only for two reasons. Either there is a point $q$ in the 
interior of the median  different from $p$ such that $u^{t^{*}}_{d_j}(q)=0$ or 
the same thing happens for the
point $q'$ that is in intersection of the median with the boundary of
$\Omega_t$. Let us examine these two cases separately. In the first case
$u^{t^{*}}_{d_j}\ge 0$ along the median but it vanishes in some intermediate
position. By symmetry it will happen in $u^{t^*}_{d_1}$ and $u^{t^*}_{d_2}$
simultaneously. Thus $u^{t^*}_{o_1}$ is an harmonic function in the yellow
region that it is positive in the boundary (and strictly positive on some
points in the boundary, for instance near $p$). Thus, by the maximal principle,
it is a strictly positive function in the interior of the yellow region. Thus
$u^{t^*}_{o_1}$ is positive in the median that bisects the yellow region. By
symmetry again $u^{t^*}_{o_3}$ is positive in the piece of the median denoted by
$o_3$ in the picture. Therefore finally $u^{t^*}_{d_1}\ge 0$ on the region
delimitated by $o_1,o_2$ and $\Omega_t$. Finally since $u^{t^*}_{d_1}$ is
harmonic it implies that it is strictly positive on the interior, i.e. on the
median $d_1$. Thus such $q$ does not exist. 

On the other hand $u^{t^*}_{d_1}$ cannot vanish on the endpoint $e$ where the
median $d_1$ meets the circle because we are assuming that $t^{*}<t_0$ and
therefore the expected lifetime near the boundary of the disk can be estimated
from below by the expected lifetime of a corona around the disk. This has an
explicit expression that has positive derivative on the boundary. Thus
$u^{t^*}_{d_1}(e)>0$. We have reached a contradiction. 

It only remains to prove that we can start the argument, i.e. that that there is
a $\delta$ such that  for a given $t<\delta$ we have that $u^t_{d_j}(x)>0$ for
any $x$ in the median joining $p$ with a vertex (excluding the barycenter).
This is the case when $t=0$. In this case we define $u^0=v$, the Green
function. In this case the gradient $v_{d_j}>0$ along the median because by the
results of \cite{LinLung} $v$ has $p$ as unique critical point in the
interior of $\Omega_0$. For very small $t$ the Green function $v$ has values in
the circles around the vertices of the triangle very close to a constant. Thus
$u_t$ can be obtained by correcting $u_t$ with an harmonic function that in the
circles almost coincides with a constant. One can check that $u_{d_j}$ is close
to $v_{d_j}$ and thus it is positive if $t$ is small enough.
\end{proof}

\end{document}